\documentclass[aps,prl,twocolumn,superscriptaddress]{revtex4-1}%
\usepackage{amsmath,amssymb,amsbsy,subfigure,hyperref,subfigure,bbm,times,txfonts,accents,float}
\usepackage{graphicx}
\usepackage{dcolumn}
\usepackage{bm}
\emergencystretch=\maxdimen
\hypersetup{colorlinks=true,linkcolor=blue,anchorcolor=blue,citecolor=blue}
\DeclareMathOperator*{\Motimes}{\text{\raisebox{0.25ex}{\scalebox{0.8}{$\bigotimes$}}}}
\begin{document}

\title{Multiparticle Entanglement Resolution Analyzer based on Quantum-Control-Assisted Uncertainty Relation}

\author{Shao-Qiang Ma}
\affiliation{%
 School of Physics, Beihang University, Beijing 100191, China
}%
\author{Xiao Zheng}
\affiliation{%
 School of Physics, Beihang University, Beijing 100191, China
}%
\author{Guo-Feng Zhang }%
\email{gf1978zhang@buaa.edu.cn}
\affiliation{%
 School of Physics, Beihang University, Beijing 100191, China
}%

\author{Heng Fan}
\affiliation{
 Beijing National Laboratory for Condensed Matter Physics, Institute of Physics, Chinese Academy of Sciences, Beijing 100190, China
}%
\affiliation{
 School of Physical Sciences, University of Chinese Academy of Sciences, Beijing 100190, China
 }
 \affiliation{
 CAS Central of Excellence in Topological Quantum Computation, Beijing 100190, China
 }
\author{Wu-Ming Liu}
\affiliation{
 Beijing National Laboratory for Condensed Matter Physics, Institute of Physics, Chinese Academy of Sciences, Beijing 100190, China
}%
\affiliation{
 School of Physical Sciences, University of Chinese Academy of Sciences, Beijing 100190, China
}
\affiliation{
 Songshan Lake Materials Laboratory , Dongguan, Guangdong 523808, China
}
\author{Leong-Chuan Kwek}
\affiliation{
 Centre for Quantum Technologies, National University of Singapore, Singapore 117543, Singapore
}%
\affiliation{
 MajuLab, CNRS-UNS-NUS-NTU International Joint Research Unit, Singapore UMI 3654, Singapore
}
\affiliation{
 National Institute of Education, Nanyang Technological University, Singapore 637616, Singapore
}

\date{\today}

\begin{abstract}
We construct a quantum-control-assisted multi-observable variance-based uncertainty relation, and the  uncertainty relation obtained indicates that we can prepare a quantum state, in which the measurement results of any observables can be predicted precisely with the help of  quantum control and  entanglement resource. The new variance-based uncertainty relation provides a concept of ``multiparticle entanglement resolution lines" for multiparticle entangled pure states, that can delineate different multiparticle entanglement classes, and thus it can be considered as an analyzer of multiparticle entanglement classes. The analyzer is used to identify different multiparticle entanglement classes  with a suitable number of incompatible measurements,  without a complete knowledge of the quantum state.
\end{abstract}
\maketitle
The fundamental difference between quantum and classical mechanics is that the measurement results of the observables are unpredictable even if all the information of the state of the quantum system is known \cite{1,2,3,4}.  This ``unpredictability" is usually expressed by the quantum uncertainty relation \cite{4D,5,6}. The uncertainty of the measurement result of an observable can be quantified by its variance and entropy, and thus these uncertainty relations are divided into the variance-based uncertainty relations \cite{2,3,7,8,9,10,11,12} and the entropic uncertainty relations \cite{4,4D,5,6,13,14,15,16,17}. Both forms of uncertainty relations arise from the preparation of a quantum state in which the measurement results of  incompatible observables cannot be predicted precisely at the same time \cite{6,8,12}. In 2010, Berta \emph{et al.} constructed an entropic uncertainty relation with the assistance of a quantum memory \cite{13}. The  uncertainty relation obtained indicates that, with the help of an entanglement resource and a quantum memory, Bob can prepare a quantum state for Alice, in which the measurement results of two given incompatible observables are predicted accurately by Bob \cite{13C,13R}. That is to say, the lower bound of the traditional uncertainty relation can be broken with the introduction of a quantum memory. Indeed, quantum-memory-assisted uncertainty relation is, essentially, a conditional uncertainty relation,  and this conditional uncertainty relation opened up new directions for a deeper understanding of quantum uncertainty relations \cite{18,19,20,21,22,23}. Yet, conditional uncertainty relations, despite possible experimental realization, remain till today largely unexplored \cite{8}.

Entanglement, a nonlocal correlation of two or more systems \cite{25,26}, is an indispensable ingredient in the quantum information science  \cite{25,221,222,223}, with ramification to  quantum simulation \cite{223} and quantum metrology \cite{0231,0232,0233,0234}. Detection of entanglement is therefore an important tool \cite{2320,241}.  In general, entanglement for two particles is easily characterized and identified \cite{25,26}, but the same question  for the multiparticle system is difficult \cite{2320,241,242,243,244}. Besides,  multiparticle entanglement belongs to different  inequivalent classes \cite{24,25,26,27,28}, and there exists no effective method to distinguish these classes when we do not capture a complete knowledge of the density matrix \cite{29,30,31,32,33}. The reconstruction of the full density matrix is usually done by the quantum state tomography \cite{34,243}, but the method becomes impossible for the states with a large number of particles \cite{243}.

One of the most important applications of the uncertainty relation is entanglement witness \cite{23,231,232,234,233}, and the advantage of the uncertainty relation over other methods is that it allows us to detect entanglement without a complete knowledge of the quantum states \cite{231,232,233,234}. Generally, the uncertainty relation can be used to identify bipartite and tripartite entanglement \cite{231,232,233,235,236,237}, but it becomes useless for four and more particles \cite{234,236}. Lots of effort has been made to overcome this problem, but none of them provides an effective solution.

In this Letter, we introduce multiparticle entanglement using variance-based uncertainty relation by constructing a multi-observable uncertainty relation through multiple quantum control systems. The new uncertainty relation shows that, even without the quantum memory, the lower bound of the traditional uncertainty relation can be broken with the help of quantum control. More importantly, based on the new uncertainty relation, we introduce the concept of ``multiparticle entanglement resolution lines" (MERLs). The MERLs present different spectral for different multiparticle entanglement classes, and thus they can be considered as a form of analyzer of multiparticle entanglement classes. Remarkably, this new analyzer can identify the different multiparticle entanglement classes even when we do not have a complete knowledge of the quantum state.

\emph{Quantum-Control-Assisted Uncertainty Relation.--- }The most famous variance-based uncertainty relation is \cite{2,3}:
\begin{align}\label{1}
\mathrm {V}\left(R\right)\mathrm {V}\left(S\right)\geq\frac{1}{4}\left|\mathrm {E}\left(\left[R,S\right]\right)\right|^2,\tag{1}
\end{align}
 where $R$ and $S$ are two incompatible observables, $\mathrm {E}([R,S])$ represents the expected value of the commutator $[R,S]$, and $\mathrm {V}(Y)$ is the variance of measurement result corresponding to the observable $Y$ with $Y\in\{R,S\}$.

In order to introduce multiparticle entanglement into variance-based uncertainty relation, we first consider the concept of the conditional variance. Traditional conditional variance is defined contingent on  a measurement and its specific result; i.e.:
\begin{align}\label{2s}
\mathrm{V}\left(Q^{S_1}|O^{S_2}:=\lambda_m\right)=\mathrm{Var}\left(Q^{S_1}, \rho^{S_1}_{\left(O^{S_2}:=\lambda_m\right)}\right) \tag{2},
\end{align}
 where $Q^{S_i}$ $(O^{S_i})$ represents the observable $Q$ $(O)$ of the subsystem $S_i$, $S_i$ stands for a subsystem of the whole system with $i\in\{1,2\}$, $O^{S_i}:=\lambda_j$ means that the result of the measurement $O$ on the subsystem $S_i$ is $\lambda_j$   with $\lambda_j$ being an eigenvalue of $O$, and $\rho_{\left(O^{S_i}:=\lambda_j\right)}$ represents the corresponding state of the whole system after the measurement with $\rho$ being the state before the measurement. $\mathrm{Var}\left(Q^{S_1}, \rho^{S_1}_{\left(O^{S_2}:=\lambda_j\right)}\right)$$=\mathrm{Tr}\left(\rho^{S_1}_{\left(O^{S_2}:=\lambda_j\right)}{Q^{S_1}}^2\right)-\left[\mathrm{Tr}\left(\rho^{S_1}_{\left(O^{S_2}:=\lambda_j\right)}Q^{S_1}\right)\right]^2$ stands for the variance of $Q^{S_1}$ on the state $\rho^{S_1}_{\left(O^{S_2}:=\lambda_j\right)}$, with $\rho^{S_1}_{\left(O^{S_2}:=\lambda_j\right)}=\mathrm{Tr}^{S_2}[\rho_{\left(O^{S_2}:=\lambda_j\right)}]$ being the partial trace of $\rho_{\left(O^{S_2}:=\lambda_j\right)}$ over the basis of the subsystem $S_2$. Thus, $\mathrm{V}\left(Q^{S_1}|O^{S_2}:=\lambda_j\right)$ represents the variance of $Q^{S_1}$ after the measurement $O^{S_2}$ has been performed and the result is obtained as $\lambda_j$.

 However, there exists a limitation in the definition, which greatly restricts the application of the conditional variance in the quantum uncertainty relation. That is, the variance of a measurement result is not reduced with the information gained from the given condition, and therefore the given condition does not reduce the uncertainty of a measurement result. For instance, considering two qubit subsystems, we have $\mathrm{V}\left(Q^{S_1}|O^{S_2}:=0\right)\geq\mathrm{V}\left(Q^{S_1}\right)$ when choosing $Q=O=\sigma_z$ and taking $\left(|0\rangle_1+|1\rangle_1\right)|0\rangle_2/2+|1\rangle_1|1\rangle_2/\sqrt{2}$ as the state of the whole system, with $|0\rangle$ and $|1\rangle$ being the eigenstates of $\sigma_z$, and  $|\cdots\rangle_i$ being the state of subsystem $S_i$.

 In order to fix this problem, we redefine the conditional variance as:
 \begin{align}\label{2ss}
\mathrm{E}\left[\mathrm{V}\left(Q^{S_1}|O^{S_2}\right)\right]=\sum_{j=1}\mathrm{P}\left(O^{S_2}:=\lambda_j\right)\mathrm{V}\left(Q^{S_1}|O^{S_2}:=\lambda_j\right) \tag{3},
\end{align}
 where $\mathrm{P}\left(O^{S_i}:=\lambda_j\right)$ represents the probability that the measurement result is $\lambda_j$ when we perform the measurement $O$ on the subsystem $S_i$. The new definition of the conditional variance is actually the expectation of $\mathrm{V}\left(Q^{S_1}|O^{S_2}:=\lambda_j\right)$ in terms of the probability $\mathrm{P}\left(O^{S_2}:=\lambda_j\right)$. Based on (\ref{2s}), $\mathrm{V}\left(Q^{S_1}|O^{S_2}:=\lambda_j\right)$  is used to quantify the variance of $Q^{S_1}$  if the measurement $O^{S_2}$ has been performed and the corresponding result is obtained as $\lambda_j$. Hence,  $\mathrm{E}\left[\mathrm{V}\left(Q^{S_1}|O^{S_2}\right)\right]$, the expectation of  $\mathrm{V}\left(Q^{S_1}|O^{S_2}:=\lambda_j\right)$ , is actually the expectation of the remaining variance of $Q^{S_1}$ when the measurement $O^{S_2}$  has been performed. That is to say, the new definition of the conditional variance  $\mathrm{E}\left[\mathrm{V}\left(Q^{S_1}|O^{S_2}\right)\right]$ is used to quantify how much variance of $Q^{S_1}$ remains after the measurement $O^{S_2}$  has been performed. Remarkably, we can obtain (please see the Section I in the Supplemental Material \cite{52}):
 \begin{align}\label{2sss}
\mathrm{E}\left[\mathrm{V}\left(Q^{S_1}|O^{S_2}\right)\right]\leq \mathrm{V}\left(Q^{S_1}\right)\tag{4},
\end{align}
which means,  under the new definition of conditional variance, the uncertainty of the measurement is reduced subject to the given condition.

\begin{figure}
\centering 
\includegraphics[height=4.5cm]{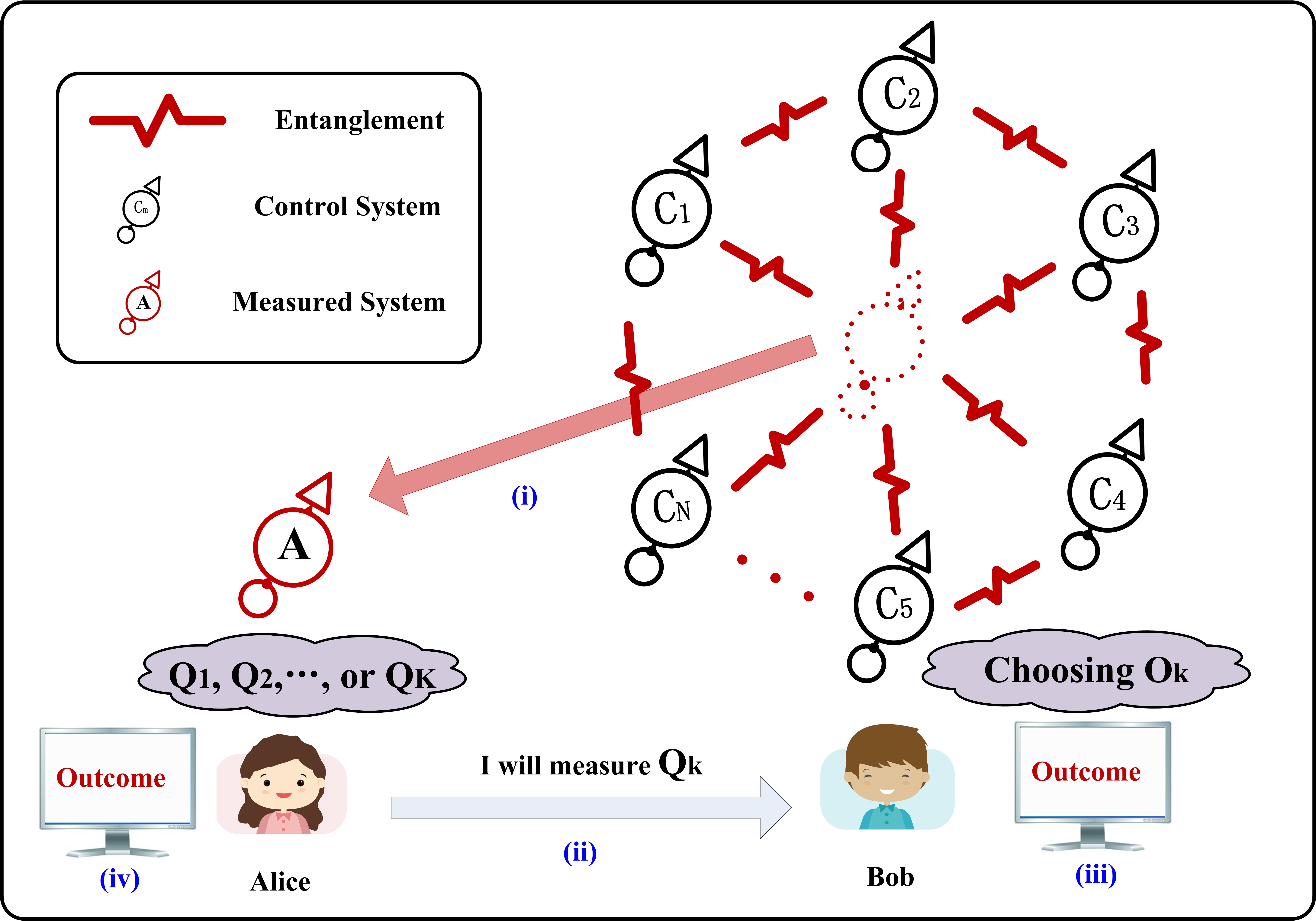}
\caption{The new uncertainty relation can be interpreted by the following uncertainty game. (i) Consider that Bob prepares $N+1$ particles $A, C_1, C_2,\cdots, C_N$, which are, in general, entangled with each other, and sends $A$ to Alice. (ii) Alice tells Bob about the measurement to be made, for instance $Q_k$, with $Q_1,Q_2,\cdots,Q_K$  being $K$ arbitrary observables and $k\in\{1,2,\cdots,K\}$. (iii)Based on the information provided by Alice and Bob's knowledge about the quantum state of the whole system, Bob chooses a appropriate measurement, for instance $O_k$,  and performs it on $C_1,C_2,\cdots,C_N$, respectively. (iv) Alice performs the measurement $Q_k$ on particle $A$. From the perspective of Bob, the uncertainty of the measurement performed by Alice is lower-bounded by the uncertainty relation (\ref{3}). The $A$ and $C_m$ are labeled as the measured system and control system, respectively, with $m\in\{1,2,\cdots,N\}$. Remarkably, the result of the measurement can be predicted precisely by Bob, when the measured system and one of control systems are in the maximum entangled state. That is to say, with the help of entanglement and quantum control, Bob can prepare a quantum state for Alice, in which the uncertainties of any observables are  arbitrarily small at the same time.} 
\label{s1}
\end{figure}

Taking advantage of the new definition of conditional variance, we  introduce multiparticle entanglement into  variance-based uncertainty relation. Consider the following game, as shown in Fig.\ref{s1}. (i) Bob prepares $N+1$ particles, denoted by $A$, $C_1, \cdots , C_{N-1}$  and $C_N$, which are entangled with each other, and sends the $A$-particle to Alice. (ii) Alice tells Bob about the measurement to be made. (iii)
Based on the information provided by Alice and Bob's knowledge about the quantum state of the whole system, Bob chooses a appropriate measurement and performs it on $C_1, C_2,\cdots,C_N$, respectively. Here we should mention that the measurements performed by Bob are not necessary the same as the measurement chosen  by Alice.
(iv) Finally, Alice performs the measurement on the particle $A$.

Based on the classical information captured by Bob, the measurements performed on the subsystems $C_1, C_2,\cdots,C_N$ are mainly used by Bob to manipulate the state of the whole system so as to minimize the uncertainty of the local measurement performed on the subsystem $A$ as much as possible. Thus, we label the particle $A$ as the measured system, the particles $C_1, C_2,\cdots,C_N$ as the control systems, and the measurements on the corresponding control systems as the quantum control. Obviously, from the perspective of Bob who possesses the control systems, the uncertainty of the measurement on the measured system is reduced with the entanglement between the measured system and control systems. In particular, by choosing a appropriate quantum control, the measurement result observed by Alice is precisely predicted by Bob when the control systems and measured system are in the maximum entangled state. The corresponding uncertainty relation can be expressed by the following inequality (please see the Section II in the Supplemental Material \cite{52}):

\begin{align}\label{3}
\sum_{k=1}^K\mathrm {E}[\mathrm {V}(Q_k^A|O_k^{C_1},\cdots,O_k^{C_N})]\geq L_{tra}-\sum_{k=1}^K\mathrm {V}[\mathrm {E}(Q_k^A|O_k^{C_1})&]\nonumber \\
-\sum_{k=1}^K\sum_{n=2}^N\mathrm {E}[\mathrm {V}(\mathrm {E}[Q_k^A|O_k^{C_n}]|O_k^{C_1},\cdots,O_k^{C_{n-1}})]&\tag{5},
\end{align}
where $Q_1,Q_2,\cdots,Q_K$ represent $K$ arbitrary incompatible observables, $O_1,O_2,\cdots,O_K$ stand for $K$ observables, and $\mathrm {E}\left[\mathrm {V}\left(Q_k^A|O_k^{C_1},\cdots,O_k^{C_N}\right)\right]$ is the conditional variance of $Q_k^A$ on the condition that we have performed the measurements $O_k^{C_1},O_k^{C_2},\cdots,O_k^{C_N}$. $\mathrm {V}\left[\mathrm {E}\left(Q_k^A|O_k^{C_1}\right)\right]$ represents the variance of $\mathrm {E}\left(Q_k^A|O_k^{C_1}:=\lambda^{(k)}_j\right)$ in terms of the possibility $\mathrm {P}\left(O_k^{C_1}:=\lambda^{(k)}_j\right)$ with $\lambda^{(k)}_j$ being an eigenvalue of $O_k$, and $\mathrm {E}[\mathrm {V}(\mathrm {E}[Q_k^A|O_k^{C_n}]|O_k^{C_1},\cdots,O_k^{C_{n-1}})]$ stands for the conditional variance of $\mathrm {E}\left(Q_k^A|O_k^{C_n}:=\lambda^{(k)}_j\right)$  on the condition that the measurements $O_k^{C_1},O_k^{C_2},\cdots,O_k^{C_{n-1}}$ have been performed \cite{54}. $L_{tra}$ represents the lower bound of the traditional sum form variance-based uncertainty relation, namely $\sum_{k = 1}^K\mathrm{V}(Q_k^A )\geq L_{tra}$. Remarkably, based on Ref.\cite{12}, the lower bound $L_{tra}$ can be exactly equal to $\sum_{k = 1}^{K}\mathrm {V}(Q_k^A)$  by introducing auxiliary operators, and the uncertainty relation (\ref{3})  will become an equality when $\sum_{k = 1}^K\mathrm{V}(Q_k^A)= L_{tra}$.

We then show that, using quantum entanglement resources, the lower bound of the traditional uncertainty relation is broken by the means of quantum control. Assuming that one of the control systems, says $C_1$, and measured system are in the maximum entangled state, and the measurements $Q_k$ and $O_k$ are the same, we deduce that  the result of measurement $Q_k^A$ is predicted precisely with the result of the measurement $O_k^{C_1}$, and thus the lower bound (\ref{3}) turns into $L_{tra}-\sum _{k=1}^{K} \mathrm {V}(O_{k}^{A})\leq0$. That is to say, the measurement results of the incompatible observables $Q_1^A,Q_2^A,\cdots,Q_K^A$ are  predicted precisely by Bob with the help of the control system. Alternatively, the lower bound  (\ref{3})  reduces to $L_{tra}$ when the entanglement between the control systems and measured system is equal to zero. In such case, the measurements on the control systems, namely the quantum control, have no effect on the measured system, and the uncertainty relation (\ref{3}) turns into the traditional sum form one.

\emph{Multiparticle Entanglement Resolution Analyzer.--- }Entanglement for two particles can be easily characterized and identified, but the same question for $M$ particles is difficult  with $M>2$. Unlike two-particle entanglement,  entanglement for $M$ particles is divided into several classes \cite{242,243,244}. Consider a fixed number $L$ with $M\geq L\geq 2$, a pure state $|\Phi\rangle$ of the whole system is $L$-separable when $|\Phi\rangle$ can be written as a tensor product of $L$ local states \cite{24}:
\begin{align}\label{4}
|\Phi\rangle=\Motimes_{l = 1}^L|\phi_l\rangle \tag{6},
\end{align}
where the local state $|\phi_l\rangle$ is the state on the subsets of the $M$ parties \cite{24}. There exists entanglement when the state is not an $L$-separable state, and the state is a genuinely multiparticle entangled state when it is not 2-separable. In general, it is difficult to delineate the multiparticle entanglement classes when we do not possess a full knowledge of the state.

We now show that the new conditional uncertainty relation can be used to identify the different classes of the multiparticle entangled pure state without a complete knowledge of the state. As shown in Fig.\ref{s1}, considering $N+1$ particles $A,C_1,\cdots,C_N$, and using $\mathcal {L}_m$ to represent the lower bound conditioned on $m$ control systems with $0\leq m\leq N$, then we have:
\begin{align}\label{5}
\mathcal {L}_0=&L_{tra}; \nonumber \\
\mathcal {L}_1=&\mathcal{L}_0 -\sum_{k = 1}^{K}\mathrm{V}\left[\mathrm {E}\left(Q^{A}_{k}|O^{C_1}_{k}\right)\right];\nonumber \\
\mathcal {L}_2=&\mathcal{L}_{1}-\sum_{k =1}^{K}\mathrm {E}\left[ \mathrm {V}\left(\mathrm {E}\left[Q^{A}_{k}| O^{C_2}_{k} \right] | O^{C_1}_{k} \right) \right] ; \nonumber \\
&\vdots \nonumber \\
\mathcal{L}_N=&\mathcal{L}_{N-1}-\sum_{k=1}^{K}\mathrm {E}\left[\mathrm{V}\left(\mathrm{E}\left[Q^{A}_{k}|O^{C_N}_{k}\right]|O^{C_1}_{k}, \ldots ,O^{C_{N-1}}_{k}\right) \right]\tag{7}.
\end{align}
The reason why the uncertainty relation can be used to detect entanglement is that the local uncertainty relation is violated whenever entanglement exists. Thus, a violation of the local uncertainty relation is considered as a sign of entanglement. Following this scheme, we deduce that the control system $C_1$ is entangled with the measured system when the traditional lower bound is broken with the introduction of $C_1$, namely $\mathcal {L}_{0}>\mathcal{L}_{1}$. Similarly, we can deduce that the state of $N+1$ particles is a genuinely  multiparticle entangled state if $\mathcal {L}_{0}>\mathcal{L}_{1}>\dots>\mathcal{L}_{N}$.

\begin{figure}
\centering 
\includegraphics[height=4.2cm]{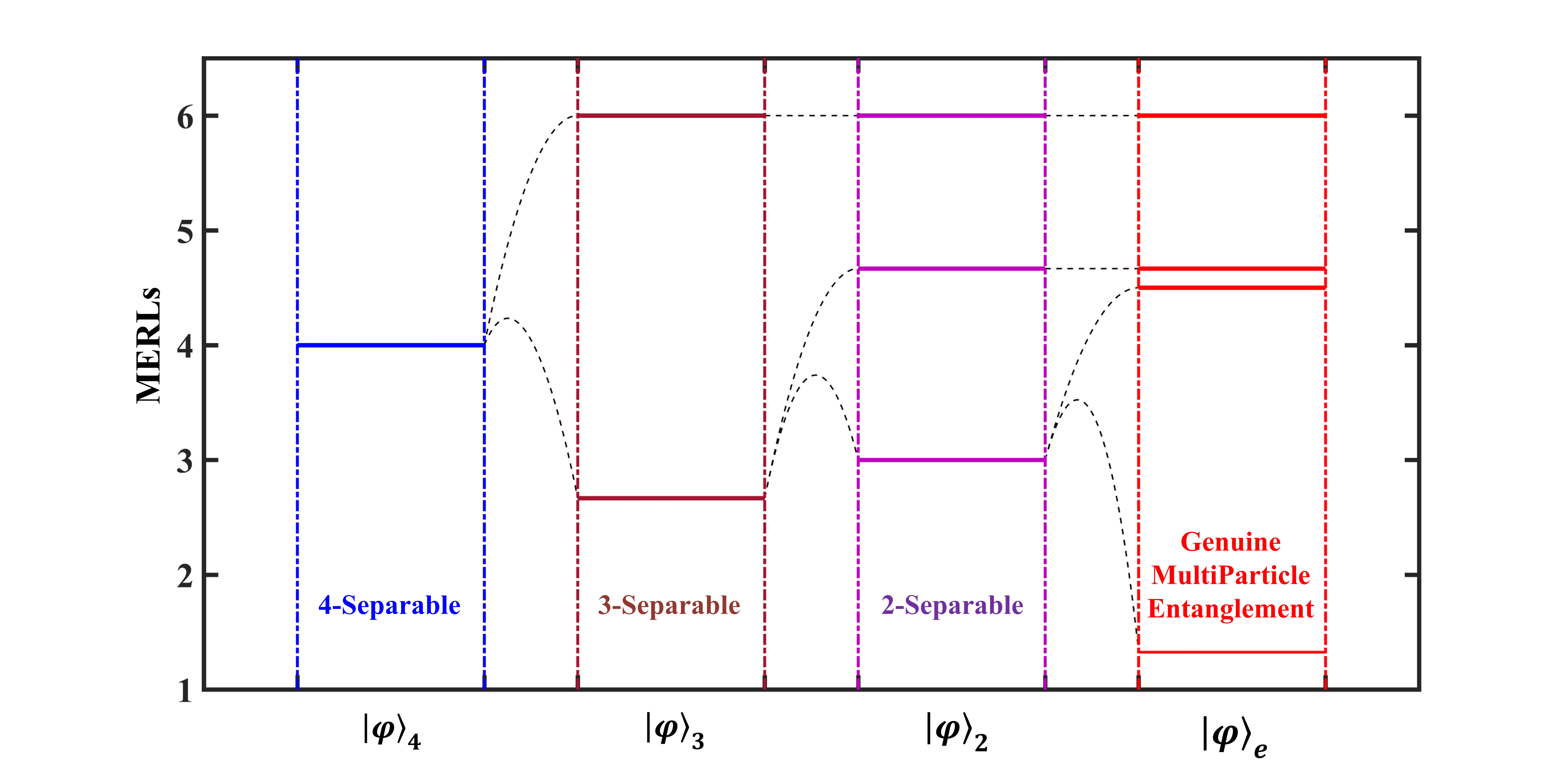}
\caption{ The MERLs of multiparticle entanglement in 4 spin-1/2 systems is presented. The subsystems are denoted by $S_1,S_2,S_3$  and $S_4$, and we take $S_1$ as the measured system and $S_2,S_3,S_4$ as the control systems. The genuinely multiparticle entangled state is taken as $|\varphi \rangle_{e}=1/\sqrt{2}(|0000\rangle+|1111\rangle)$, the 2-separable state is taken as $|\varphi\rangle_2 =1/\sqrt{2}(|000\rangle+|111\rangle)\otimes|0\rangle$, the 3-separable state is taken as $|\varphi\rangle_{3}=1/\sqrt{2}(|00\rangle+|11\rangle)\otimes|0\rangle\otimes|0\rangle$, and the 4-separable state is taken as $|\varphi\rangle_{4}=|0\rangle\otimes|0\rangle\otimes|0\rangle\otimes|0\rangle$, where $|0\rangle$ and $|1\rangle$ are the eigenstates of $\sigma_z$. We take $K=4$, and the incompatible observables are taken as $Q_1=O_1=\sigma_x, Q_2=O_2=\sigma_y, Q_3=O_3=\sigma_z$, and $Q_4=O_4=\sigma_x+\sigma_y+\sigma_z$. The traditional lower bound $L_{tra}$ is taken as the lower bound of uncertainty equality, namely $L_{tra}=\sum_{k=1}^K\mathrm {V}(Q^A_k)$, and then the MERLs of $|\varphi\rangle_e , |\varphi\rangle_2 ,|\varphi\rangle_3 $ and $|\varphi\rangle_4 $ is obtained. We see that the genuinely multiparticle entangled state is identified by the split MERLs. Here the black dashed lines represent the splitting process of the MERLs \cite{56}.
} 
\label{s2}
\end{figure}

Here, the lower bounds $\left\{\mathcal {L}_{0},\mathcal{L}_{1}, \dots, \mathcal{L}_{N}\right\}$ are named as ``multiparticle entanglement resolution lines" (MERLs), and in general the MERLs are coincide with each other, because we have $\mathcal {L}_{0}\geq\mathcal{L}_{1}\geq\dots\geq\mathcal{L}_{N}$ and  at least one equal sign is established for non-genuinely multiparticle entangled state. That is to say, the MERLs completely split from each other only for a genuinely multiparticle entangled state. Thus, the split MERLs is used to identify genuinely multiparticle entangled pure state; i.e., a pure state is genuinely multiparticle entangled when the number of the split MERLs is exactly equal to the number of the subsystems, as shown in Fig.\ref{s2}.  Moreover, it can be seen from Fig.\ref{s2} that the split MERLs are also used to identify the $L$-separable state. We deduce that a state is $(N+2-m)$-separable, $(N+1-m)$-separable, $\cdots$ , $2$-separable or genuinely multiparticle entangled when the $m$ split MERLs have been detected for the system which consists of $N+1$ subsystems. Here, we should mention that the conclusions obtained above only apply for the case of pure states, and the related proofs are presented in the Section III of the Supplemental Material \cite{52}).

In general, the traditional lower bound $L_{tra}$ is taken as $\sum_{k = 1}^{K}\mathrm {V}(Q_k^A)$ , and therefore the uncertainty relation (\ref{3}) becomes an equality. Then, the MERLs are rewritten as $\mathcal {L}_m=\sum_{k=1}^K\mathrm {E}[\mathrm {V}(Q_k^A|O_k^{C_1},\cdots,O_k^{C_m})]$  for $m\geq1$ and $\sum_{k = 1}^{K}\mathrm {V}(Q_k^A)$ for $m=0$. Based on the new definition of the conditional variance, we can see that the MERLs are obtained only through some suitable incompatible measurements, even without a complete knowledge of the quantum state.
\begin{figure}
\centering 
\includegraphics[height=4.2cm]{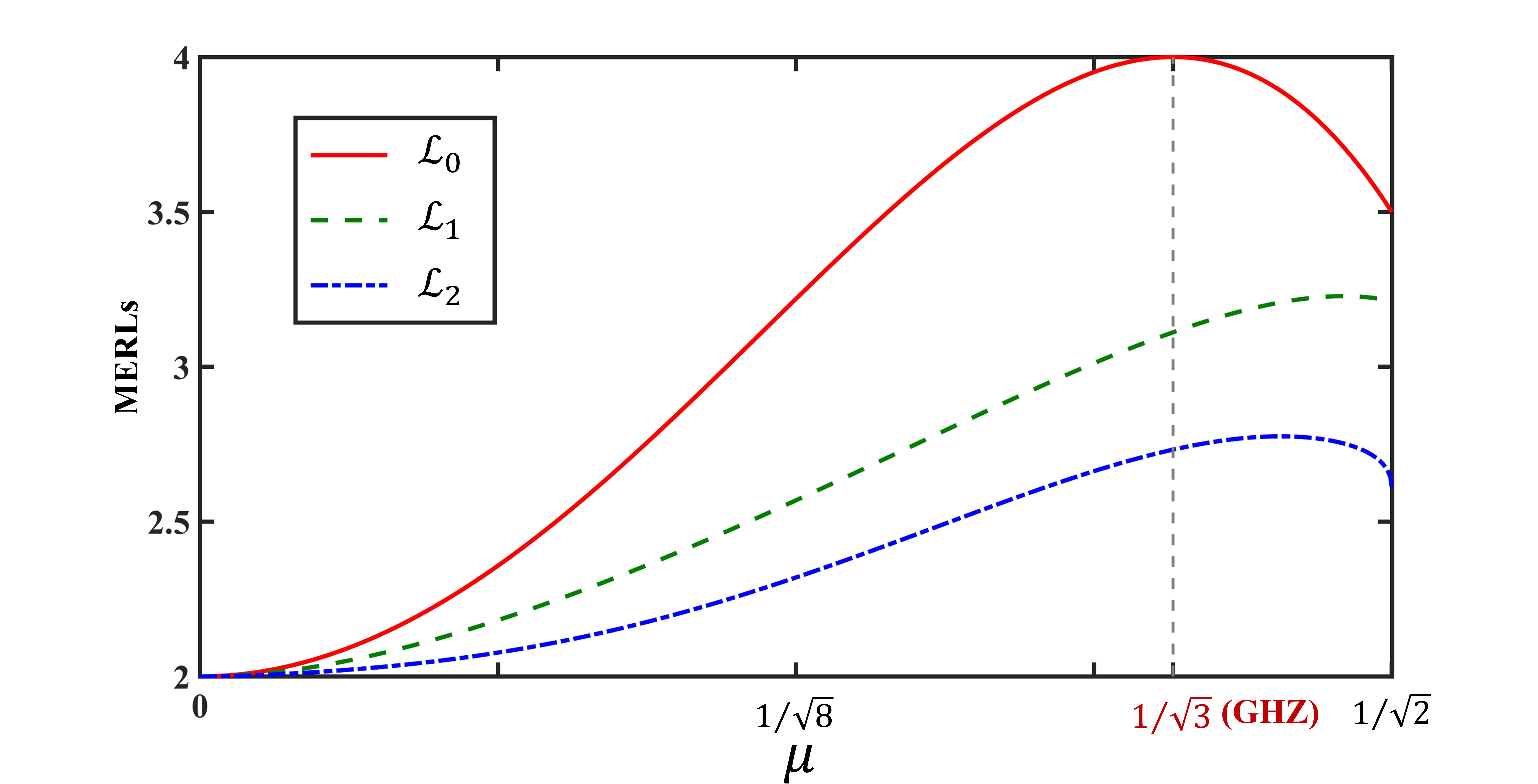}
\caption{Here the incompatible observables, $Q_1=O_1=J_x$, $Q_2=O_2=J_y$, $Q_3=O_3=J_z$ and $Q_4=O_4=J_x+J_y+J_z$ \cite{59} are chosen to verify genuine multiparticle entanglement, where the observables calculated on the eigenvectors $\left\{|2\rangle_1,|-1\rangle_1,|3\rangle_1\right\}$ for the first photon, the eigenvectors $\left\{|0\rangle_2,|-1\rangle_2,|1\rangle_2\right\}$ for the second photon, and $\left\{|0\rangle_3,|-1\rangle_3,|1\rangle_3\right\}$ for the third photon, respectively. We can see that the GHZ state is actually identified by the MERLs.
} 
\label{s3}
\end{figure}

\emph{GHZ State Witness.--- }Here, an example will be presented to demonstrate the superiority of the MERLs. In many experiments, due to technical limitation, the research on multiparticle entanglement has been confined mainly on qubits. However, entanglement beyond qubit, such as multiparticle entanglement based on the qutrit system, provides  more information than the  entangled qubit state, and thus a lot of work has been done to prepare the genuinely multiparticle entangled state beyond qubit system in experiment. In 2016, using the orbital angular momentum of the photon, Malik \emph{et al.} prepared a high dimension multiparticle entangled state \cite{57}, where two photons reside in a three-dimension space and the third one lives in a two-dimension space (denoted by (3,3,2)-type multiparticle entanglement). In 2018, they create a real three-particle GHZ state entangled in (3,3,3) way \cite{58}, which reads:
\begin{align}\label{6}
\sqrt{1-2\mu^2}|2\rangle_1|0\rangle_2|0\rangle_3+\mu|-1\rangle_1|-1\rangle_2|-1\rangle_3-\mu|3\rangle_1|1\rangle_2|1\rangle_3 \nonumber, 
\end{align}
where $|\cdots\rangle_h$ stands the orbital angular momentum quanta of the $h-th$ photon, $\mu$ changes from 0 to $1/\sqrt{2}$ by adjusting the relevant experimental parameters, and the state becomes a real GHZ state when $\mu=1/\sqrt{3}$.

With the success in the preparation of high dimension multiparticle entangled state in experiment, another problem arises. That is ``how does one verify these multiparticle entangled states experimentally". Malik  \emph{et al.} certify the prepared multiparticle entangled state by measuring the fidelity of the prepared state to the ideal GHZ state, and they show that the prepared state possesses genuine multiparticle entanglement when the fidelity is less than a certain threshold value \cite{58}. Indeed, this method still depends on the tomographic reconstruction of the quantum state, and thus the number of measurements increases dramatically with the increase in the dimension of the subsystem. For instance, as mentioned in Refs.\cite{57} and \cite{58}, the verification of the (3,3,2)-type and (3,3,3)-type genuine entanglement needs 162 and 219 total measurements, respectively. We note that, based on our analyzer, the prepared GHZ state is verified with only polynomial number of incompatible measurements, which does not change with the increase of the dimension of the subsystem , as shown in Fig.\ref{s3}.

\emph{Conclusion.--- }In this Letter, we construct a multi-observable variance-based conditional uncertainty relation by introducing multiparticle entanglement and several quantum control systems. The  uncertainty relation obtained indicates that we can prepare a quantum state, in which the measurement results of any observables are predicted precisely at the same time, when the prepared system and the control systems are in  maximum entangled state. Multiparticle entanglement is divided into several classes, and, in general, it is difficult to characterize them when we cannot obtain a full knowledge of the quantum state. Here, using the lower bounds of the new conditional uncertainty relation, we introduce the concept of ``multiparticle entanglement resolution lines" for multiparticle entangled pure states. The ``multiparticle entanglement resolution lines"  present different spectral for different multiparticle entanglement classes, and these lines are obtained only using some incompatible measurements. Finally, the ``multiparticle entanglement resolution lines" can be used to identify multiparticle entangled pure state, even without a complete knowledge of the quantum state.

This work was supported by the NSFC under Grants No. 11574022, No. 61775242, No. 11434015, No. 61227902, No. 61835013; the National Key R$\&$D Program of China under Grants No. 2016YFA0301500; and the Strategic Priority Research Program of the Chinese Academy of Sciences under Grants No. XDB01020300 and No. XDB21030300, and L. C. Kwek was supported by the National Research Foundation Singapore and the Ministry of Education  Singapore.

S. Q. M., X. Z. and G. F. Z. contributed equally to this work.


\begin{thebibliography}{99}
\bibitem{56}The splitting process between  $|\varphi\rangle$ and $|\psi\rangle$ is demonstrated by the evolution of the MERLs for the state $\epsilon|\varphi\rangle\langle\varphi|+(1-\epsilon)|\psi\rangle\langle\psi|$ with respect to $\epsilon$, with $0\leq\epsilon\leq1$ and $|\varphi\rangle,|\psi\rangle\in\{|\varphi\rangle_e,|\varphi\rangle_2,|\varphi\rangle_3,|\varphi\rangle_4\}$.
\bibitem{59}Here,$J_x$, $J_y$, and $J_z$ are qutrit spin observables.
\end{thebibliography}
\end{document}